\documentclass[prb,twocolumn,showpacs,floatfix,preprintnumbers,amsmath, amssymb,superscriptaddress]{revtex4}

\usepackage{graphicx}
\usepackage{amsmath}
\usepackage{amssymb}
\usepackage{color}
\usepackage{dcolumn}
\usepackage{epsfig}
\usepackage{bm}
\usepackage[urlcolor=blue]{hyperref}
\hypersetup{backref, colorlinks=true, linkcolor=blue, citecolor=blue}

\begin{document}

\title{Structural and bonding character of potassium-doped $p$-terphenyl superconductors}

\author{Guo-Hua Zhong}
\affiliation{Shenzhen Institutes of Advanced Technology, Chinese Academy of Sciences, Shenzhen 518055, China}
\affiliation{Beijing Computational Science Research Center, Beijing 100193, China}

\author{Xiao-Hui Wang}
\affiliation{Beijing Computational Science Research Center, Beijing 100193, China}

\author{Ren-Shu Wang}
\affiliation{Center for High Pressure Science and Technology Advanced Research, Shanghai 201203, China}

\author{Jia-Xing Han}
\affiliation{Beijing Computational Science Research Center, Beijing 100193, China}

\author{Chao Zhang}
\affiliation{Department of Physics, Yantai University, Yantai, 264005, China}

\author{Xiao-Jia Chen}
\email{xjchen@hpstar.ac.cn}
\affiliation{Center for High Pressure Science and Technology Advanced Research, Shanghai 201203, China}

\author{Hai-Qing Lin}
\email{haiqing0@csrc.ac.cn}
\affiliation{Beijing Computational Science Research Center, Beijing 100193, China}

\date{\today}

\begin{abstract}
Recently, there is a series of reports by Wang \emph{et al.} on the superconductivity in K-doped \emph{p}-terphenyl (K$_x$C$_{18}$H$_{14}$) with the transition temperatures range from 7 to 123 Kelvin. Identifying the structural and bonding character is the key to understand the superconducting phases and the related properties. Therefore we carried out an extensive study on the crystal structures with different doping levels and investigate the thermodynamic stability, structural, electronic, and magnetic properties by the first-principles calculations. Our calculated structures capture most features of the experimentally observed X-ray diffraction patterns. The K doping concentration is constrained to within the range of 2 and 3. The obtained formation energy indicates that the system at $x=2.5$ is more stable. The strong ionic bonding interaction is found in between K atoms and organic molecules. The charge transfer accounts for the metallic feature of the doped materials. For a small amount of charge transferred, the tilting force between the two successive benzenes drives the system to stabilize at the antiferromagnetic ground state, while the system exhibits non-magnetic behavior with increasing charge transfer. The multiformity of band structures near the Fermi level indicates that the driving force for superconductivity is complicated.
\end{abstract}


\maketitle
\section{Introduction}
Superconducting materials have been a hot research topic in physics and materials science due to their important application values in energy, information, quantum devices and other advanced technologies. Researchers have been devoted to the design and synthesis of new superconducting materials, the understanding of superconducting mechanism and the exploration of superconductors with higher transition temperature ($T_c$). In 2010, the superconductivity of $T_c\sim 18$ Kelvin was discovered in potassium (K) metal doped picene\cite{ref1}, which opens a new avenue in the quest for organic hydrocarbon superconductors. Subsequently, the superconductivity has also observed in alkali, alkaline-earth metals and rare-earth elements doped phenanthrene\cite{ref2,ref3,ref4,ref5}, chrysene\cite{ref5}, phenacene\cite{ref6}, coronene\cite{ref7}, phenacene\cite{ref6} and 1,2;8,9-dibenzopentacene\cite{ref8}, respectively. Especially, using density functional theory and Eliashberg's theory of superconductivity, we have predicted K-doped benzene (K$_2$C$_6$H$_6$) to be superconductive with the $T_c$ around 6.2 Kelvin, and proposed that all hydrocarbons should show $T_c$ in a similar temperature range of $5 < T_c < 7$ Kelvin under conventional conditions\cite{ref9}. The results greatly increase the understanding of the superconductivity of polycyclic aromatic hydrocarbons (PAHs).

So far the highest $T_c$ reached is 33.1 Kelvin in PAH superconductors, which was obtained in K-doped 1,2;8,9-dibenzopentacene\cite{ref8}. However, previous theoretical studies showed that PAH superconductors usually exhibit the antiferromagnetic (AFM) ground state\cite{ref10,ref11,ref12,ref13} and exist the strong electronic correlation effects\cite{ref13,ref14,ref15,ref16,ref17,ref18}, indicating the superconducting mechanism is complicated in aromatic hydrocarbons. Our theoretical predictions have argued that electronic correlations\cite{ref9} or pressure\cite{ref19} could enhance high $T_c$ in aromatic hydrocarbons. Hence, organic based compounds are candidates of high temperature or room temperature superconductors, since the interaction of electrons with much higher excitation energy than the phonon energy can result in a substantially higher $T_{c}$ in these low dimensional materials\cite{ref20,ref21}. Thus, the higher $T_{c}$ can be expected in this kind of aromatic hydrocarbons.

Recently,  \emph{p}-terphenyl (C$_{18}$H$_{14}$), a hydrocarbon compound containing three benzene rings connected by C-C bonding, draws a lot of attention. Different from the feature of sharing armchair edge in PAHs such as phenanthrene, chrysene, picene, coronene and 1,2;8,9-dibenzopentacene, benzene rings are connected by the single C-C bond in the \emph{p}-terphenyl molecule. Namely, \emph{p}-terphenyl is a non-polycyclic aromatic hydrocarbon. With regard to the study of superconductivity of this system, Wang \emph{et al.} claimed the superconductivity of 7.2 Kelvin in the synthesized K$_x$C$_{18}$H$_{14}$\cite{ref22}. Later, Wang \emph{et al.} claimed to observe a higher $T_{c}$ of 43 Kelvin\cite{ref23} and even 123 Kelvin\cite{ref24} in the K-doped \emph{p}-terphenyl. They also ruled out the possibilities of the formation of K-doped C$_{60}$ and graphite as suggested previously \cite{ref25}. By using high resolution photoemission spectroscopy on potassium surface-doped \emph{p}-terphenyl crystals, Li \emph{et al.} presented the spectroscopic evidence for pairing gaps at the surfaces of these materials, with the gaps persisting to 60 K or above\cite{ref26}. Furthermore, Liu \emph{et al.} observed the superconductivity like transition at about 125 Kelvin in their fabricated potassium doped \emph{p}-terphenyl with the help of magnetization measurements\cite{ref27}. These subsequent works greatly promote the study to superconductivity of the material. However, the doping level and superconducting phases with different $T_{c}$ have not been determined yet, not to mention the desired pairing mechanism. So identifying the structures and chemical bonds is crucial to the understanding of the superconducting phases as well as superconductivity, which is the focus of the current study.

\emph{p}-Terphenyl has twisting degrees of freedom around the long molecular axis, which will result in complicated situations when K atoms are doped into \emph{p}-terphenyl crystal. It has been well known that crystalized \emph{p}-terphenyl exhibits two phases with the variation of temperature of $P2_1/c$ symmetry at room temperature and $P\bar{1}$ symmetry below 193 Kelvin\cite{ref28,ref29}. In addition, the pressure also drives a transition from $C_2$ to $D_{2h}$ symmetry around 1.3 GPa\cite{ref30}. In this work, we will investigate the K doping effects on structural, electronic, and magnetic properties in \emph{p}-terphenyl at ambient pressure by the first-principles calculations. The obtained structures  will be compared with experiments. The details for the structural and bonding features will be provided. These results are helpful for the determination of the doping level and thermaldynamically stable phases as well as the understanding of the charge transfer process and mechanism in these newly discovered superconductors.

\section{Method}
To study the structural and electronic properties of K$_x$C$_{18}$H$_{14}$, we employed the Vienna \emph{ab} initio simulation package (VASP) \cite{ref31,ref32} based on the projector augmented wave method. For the plane-wave basis-set expansion, an energy cutoff of 600 eV was adopted. The Monkhorst-Pack \emph{k}-point grids are generated according to the specified \emph{k}-point separation of 0.02 {\AA}$^{-1}$ and the convergence thresholds are set as $10^{-6}$ eV in energy and $10^{-3}$ eV/{\AA} in force. The generalized gradient form (GGA) of the exchange-correlation functional (Perdew-Burke-Ernzerh of 96, PBE) was adopted\cite{ref33}. And considering the non-local interaction, we has added the correction of van der Waals (vdW) in version of vdW-DF2 in this calculation\cite{ref34}. The necessity of vdW-DF2 functional has been confirmed by our previous studies\cite{ref35,ref36}.

\section{Results}
\subsection{Determining functional}

\begin{table}
\caption{\label{tab:table1} The optimized crystal lattice constants $a$, $b$, $c$, the angle $\beta$ between two axes and the volume $V$ of pristine C$_{18}$H$_{14}$ comparing with the experimental values.}
\begin{ruledtabular}
\begin{tabular}{cccccc}
Method                &$a$ ({\AA}) & $b$ ({\AA}) & $c$ ({\AA}) & $\beta$ ($^{\circ}$) & $V$ ({\AA}$^{3}$) \\ \hline
Expt.\footnotemark[1] & 13.613     & 5.613       & 8.106       & 92                   & 619               \\
Expt.\footnotemark[2] & 13.621     & 5.613       & 8.116       & 92.01                & 621.1             \\
Expt.\footnotemark[3] & 13.59      & 5.59        & 8.08        & 92.9                 & 613.4             \\
Expt.\footnotemark[4] & 13.58      & 5.58        & 8.02        & 92.10                & 607.3             \\
Expt.\footnotemark[5] & 13.55      & 5.63        & 8.15        & 92.60                & 621.1             \\
vdW-DF2               & 13.569     & 5.542       & 7.911       & 92.8                 & 594.2             \\
LDA                   & 13.215     & 5.292       & 7.443       & 93.57                & 519.6             \\
GGA                   & 13.596     & 6.283       & 8.574       & 88.89                & 732.2             \\
\end{tabular}
\end{ruledtabular}
\footnotetext[1]{Ref.~\onlinecite{ref37}.}
\footnotetext[2]{Ref.~\onlinecite{ref38}.}
\footnotetext[3]{Ref.~\onlinecite{ref39}.}
\footnotetext[4]{Ref.~\onlinecite{ref40}.}
\footnotetext[5]{Ref.~\onlinecite{ref41}.}
\end{table}

Starting from the pristine C$_{18}$H$_{14}$, we firstly optimized the crystal lattice parameters of solid C$_{18}$H$_{14}$ with the $P2_1/c$ symmetry to test the feasibility of vdW-DF2 functional. The obtained crystal lattice parameters from vdW-DF2 functional are $a=13.569$ {\AA}, $b=5.542$ {\AA}, $c=7.911$ {\AA} and the angle $\beta=92.8^{\circ}$, respectively. Seen from the crystal lattice parameters listed in Table I, the vdW-DF2 functional products the lattice constants are in good agreement with experimental ones, though the small distinguish exists among those reported experimental observations\cite{ref37,ref38,ref39,ref40,ref41}. The error is respectively controlled within 2.5\% for the lattice constant and 4\% for the volume. But GGA and local density approximation (LDA)\cite{ref42} functional respectively extremely overestimates and underestimates the lattice constants (unit volume). The result indicates that the non local interaction should not be ignored in this aromatic hydrocarbon. As shown in Fig. 1, the calculated XRD spectrum of solid C$_{18}$H$_{14}$ shown in Fig. 1 fits the experiment done by Wang \emph{et al.}\cite{ref22}. Additionally, the average value of carbon-carbon bond lengths within the rings is about 1.40 {\AA} and those between rings are about 1.49 {\AA}, which are also consistent with experimental values\cite{ref29}.

\subsection{Structure of K$_x$C$_{18}$H$_{14}$}

\begin{figure}
\includegraphics[width=\columnwidth]{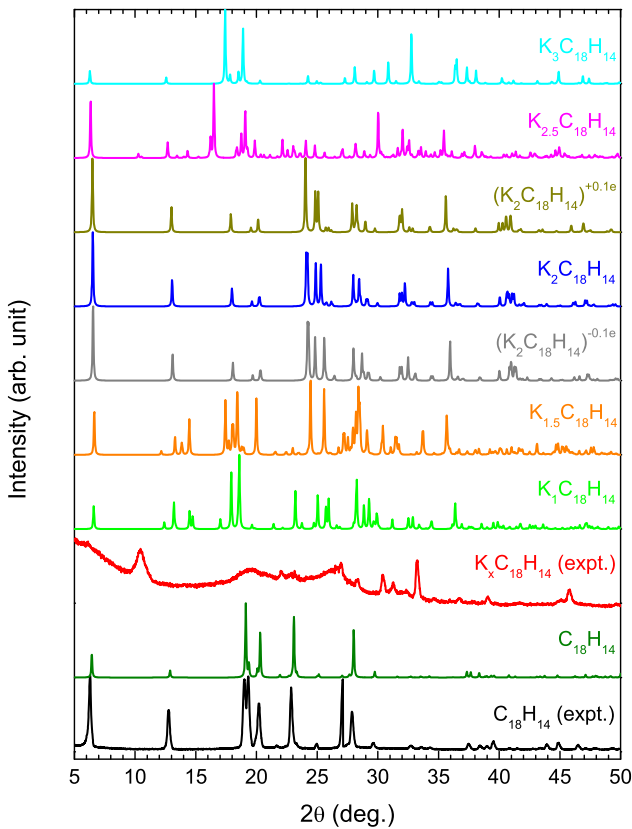}
\caption{(Color online) Calculated XRD spectra of pristine and K-doped C$_{18}$H$_{14}$ comparing with experiments\cite{ref22}. The experimental XRD patterns of K-doped C$_{18}$H$_{14}$ were taken from the sample with $T_c=7.2$ Kelvin. All XRD data was collected by using the incident wavelength $\lambda=1.5406$ {\AA}.}
\end{figure}

Within the framework of the vdW-DF2 functional, we simulated K-doped \emph{p}-terphenyl, and firstly considered  three possible concentrations of $x=$ 1, 2 and 3 in K$_x$C$_{18}$H$_{14}$. The total energy calculation indicates that the doping results in the phase transition of crystal structure. Both K$_1$C$_{18}$H$_{14}$ and K$_2$C$_{18}$H$_{14}$ are stabilized at $P2_1$ symmetry instead of $P2_1/c$. But K$_3$C$_{18}$H$_{14}$ can exist in form of $P2_1/c$ symmetry. The crystal lattice parameters at three doping levels are summarized in Table II. We find that the variation of crystal lattice parameters is complicated after doping. The intercalation of K atoms leads to the obvious expansion in $b$ direction and the big contraction in $c$ direction in K$_1$C$_{18}$H$_{14}$ and K$_2$C$_{18}$H$_{14}$. K$_3$C$_{18}$H$_{14}$ is abnormal since it expands in $a$ and $c$ directions while contracts in the $b$ direction. However, the doping makes the system volume increase. Figure 2 clearly shows the characteristics of K$_x$C$_{18}$H$_{14}$ viewed from different directions. For K$_1$C$_{18}$H$_{14}$, as shown in Fig. 2(a), the herringbone structure is formed and K atom is between two organic molecular layers when viewing along the $a$ direction, similar to K-doped picene\cite{ref1} and phenanthrene\cite{ref2}. There is a tilted angle $\delta$ between two successive benzene rings with respect to each other, and it reaches the extent of $9.5^{\circ}-10.6^{\circ}$. And viewing from the $c$ direction, K atom is localized on the C-C bond connected benzene ring-1 with ring-2, and closer to ring-1. Scanning the fine bonding feature, we have found that the C-C bonds linked two benzene rings (hereafter call them as bridge bonds) are shortened to 1.46 {\AA} from 1.49 {\AA} while the C-C bonds near the bridge bond within the rings become long when one K atom is doped for every organic molecule. The variations of C-C bonds indicate that the charge is local near K atom instead of uniform distribution when transferring to $\pi$ orbital from K atom.

\begin{table*}
\caption{\label{tab:table2} The optimized crystal lattice constants $a$, $b$, $c$, the angle $\beta$ between two axes, the tilted angle $\delta$ between two successive benzene rings with respect to each other and the volume $V$ for every doping level.}
\begin{ruledtabular}
\begin{tabular}{cccccccccc}
System                & Space-group & $a$ ({\AA}) & $b$ ({\AA}) & $c$ ({\AA}) & $\alpha$ ($^{\circ}$) & $\beta$ ($^{\circ}$) & $\gamma$ ($^{\circ}$) & $\delta$ ($^{\circ}$) & $V$ ({\AA}$^{3}$) \\ \hline
K$_1$C$_{18}$H$_{14}$ & $P2_1$      & 13.500      & 6.863       & 7.181       & 90                    & 83.12                & 90                    & 9.5-10.6              & 660.6             \\
K$_{1.5}$C$_{18}$H$_{14}$ & $P1$    & 7.110       & 7.274       & 14.743      & 89.76                 & 64.56                & 90.02                 & 17.2-20.3             & 688.6             \\
K$_2$C$_{18}$H$_{14}$ & $P2_1$      & 14.001      & 7.032       & 7.151       & 90                    & 75.54                & 90                    & 0.2-0.6               & 681.8             \\
K$_{2.5}$C$_{18}$H$_{14}$ & $P1$    & 14.585      & 6.278       & 9.058       & 99.25                 & 73.54                & 96.77                 & 1.1-2.5               & 782.7             \\
K$_3$C$_{18}$H$_{14}$ & $P2_1/c$    & 14.503      & 5.463       & 10.243      & 90                    & 76.13                & 90                    & 0                     & 787.8             \\
\end{tabular}
\end{ruledtabular}
\end{table*}

For K$_2$C$_{18}$H$_{14}$, the herringbone structure is still existent as shown in Fig. 2(b). However, three benzene rings in organic molecule are almost coplanar. The tilted angle $\delta$ is only $0.2^{\circ}-0.6^{\circ}$. Along the $b$ direction, two K atoms are between two organic molecular layers, and viewing from the $c$ direction, K atom respectively lies on the bridge bond. In K$_2$C$_{18}$H$_{14}$, the bridge bonds further shorten to 1.44 {\AA} and the C-C bonds near the bridge bond within the rings become longer. The distinguish of C-C bond lengths within the same ring becomes big which implies a stronger molecular distortion in K$_2$C$_{18}$H$_{14}$. Doping three K atoms into solid \emph{p}-terphenyl, K$_3$C$_{18}$H$_{14}$, the system keeps the $P2_1/c$ symmetry, but great changes have taken place in the structure as shown in Fig. 2(c). All organic molecules rotate and form a visible two-dimensional layered structure. Along the $b$ direction, three K atoms are all between two organic molecular layers, but viewing from the $a$ direction, K atoms and \emph{p}-terphenyl molecules are coplanar. Viewing from the $b$ direction, the K atom is respectively on the center of benzene ring. The varying tendency of C-C bond lengths in K$_3$C$_{18}$H$_{14}$ is similar to K$_1$C$_{18}$H$_{14}$ and K$_2$C$_{18}$H$_{14}$. The bridge bonds further shorten to 1.43 {\AA}.

\begin{figure}
\includegraphics[width=\columnwidth]{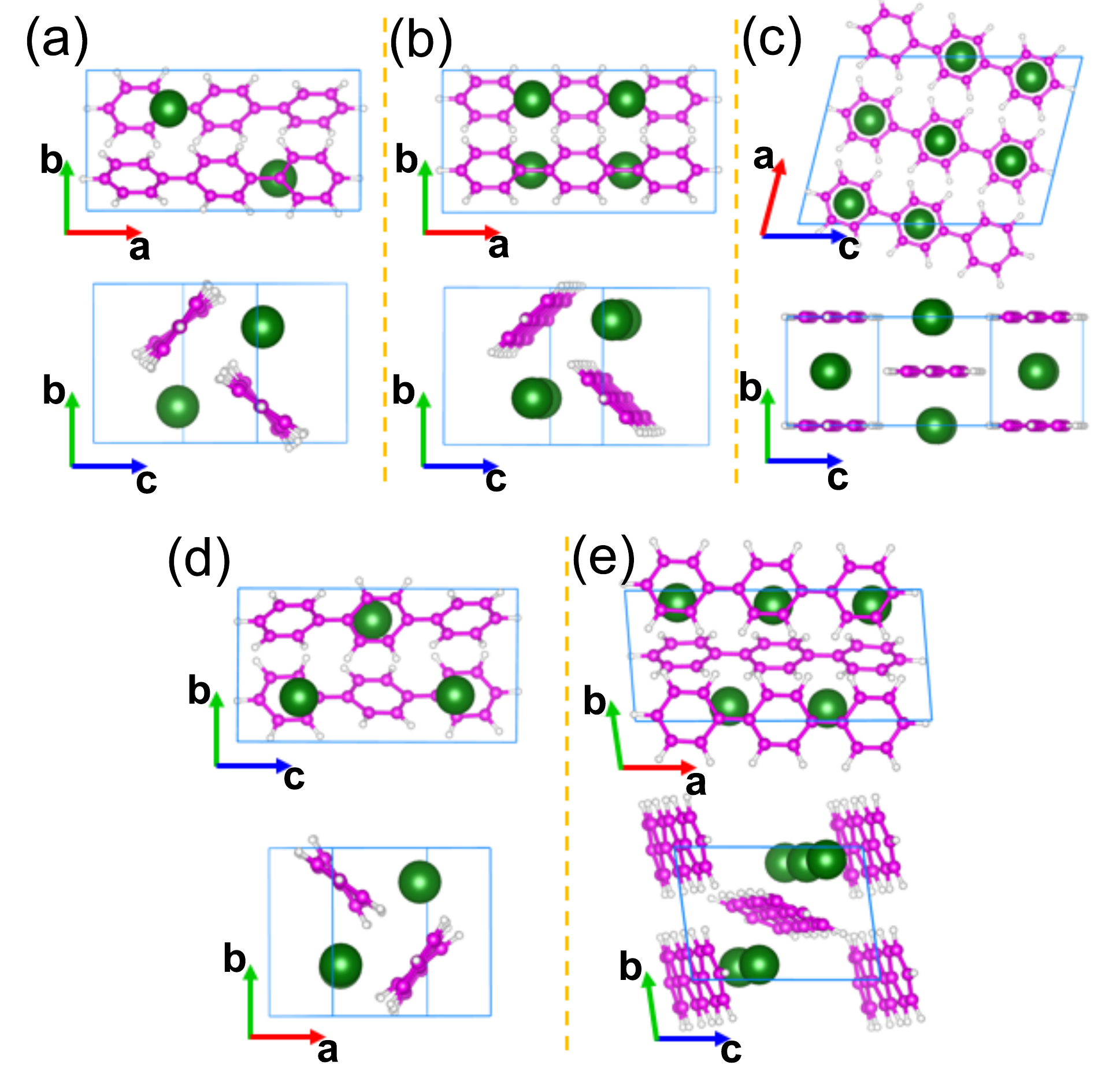}
\caption{(Color online) Optimized structures of K$_x$C$_{18}$H$_{14}$ viewing from different directions. Pink, white and green balls represent C, H and K atoms, respectively. (a) K$_1$C$_{18}$H$_{14}$, (b) K$_2$C$_{18}$H$_{14}$, (c) K$_3$C$_{18}$H$_{14}$, (d) K$_{1.5}$C$_{18}$H$_{14}$ and (e) K$_{2.5}$C$_{18}$H$_{14}$.}
\end{figure}

To identify the observed superconducting phase in experiment, we have analyzed the XRD result. Compared the experimental XRD between C$_{18}$H$_{14}$ and K$_x$C$_{18}$H$_{14}$ with 7.2 Kelvin superconductivity, the abnormal peaks around $2\theta=10.4^{\circ}$ and $2\theta=33.2^{\circ}$ imply the formation of new phase and a possible phase transition after doping. Additionally, the wide XRD peaks indicate that the crystallinity of the sample is not very good so that mixture phase can exist. Our calculated XRD of K$_1$C$_{18}$H$_{14}$, K$_2$C$_{18}$H$_{14}$ and K$_3$C$_{18}$H$_{14}$ are presented in Fig. 1. Unfortunately, we find that none of three structures is completely consistent with the experiment in XRD spectrum. Although the XRD of K$_2$C$_{18}$H$_{14}$ or K$_3$C$_{18}$H$_{14}$ is closer to the experimental result, the XRD peak around $2\theta=10.4^{\circ}$ has not observed in these two pure phases. Thus we further investigated the possibility of other doped phases. On the one hand, we added a small charge fluctuation into K$_2$C$_{18}$H$_{14}$ to form defect state, such as (K$_2$C$_{18}$H$_{14}$)$^{+0.1e}$ and (K$_2$C$_{18}$H$_{14}$)$^{-0.1e}$. However, except for the small shift of peaks, this did not change the XRD spectrum in nature as shown in Fig. 1. On the other hand, we considered the fractional doping concentrations, such as K$_{1.5}$C$_{18}$H$_{14}$ or K$_{2.5}$C$_{18}$H$_{14}$, which can also understood as a mixed phase. In experiment, K$_{1.5}$C$_{18}$H$_{14}$ may be obtained by mixing K$_1$C$_{18}$H$_{14}$ and K$_2$C$_{18}$H$_{14}$, while K$_{2.5}$C$_{18}$H$_{14}$ can be formed by mixing K$_2$C$_{18}$H$_{14}$ and K$_3$C$_{18}$H$_{14}$. By subtracting and adding K atoms from/into K$_2$C$_{18}$H$_{14}$, we have simply simulated the fractional doping levels in this calculation. Table II and Figure 2 present the optimized crystal lattice parameters and geometrical configurations, respectively. Both K$_{1.5}$C$_{18}$H$_{14}$ and K$_{2.5}$C$_{18}$H$_{14}$ are stabilized at $P1$ symmetry. The volumes of unit cell of doping cases of $x=$ 1, 1.5, 2, 2.5 and 3 does not satisfy the monotonically increasing trend due to the existence of phase transitions during the doping. As shown in Fig. 2(d), K$_{1.5}$C$_{18}$H$_{14}$ is similar to K$_1$C$_{18}$H$_{14}$. There are a visible herringbone feature and a big tilted angle between two successive benzene rings with respect to each other. The asymmetric distribution of K atoms in K$_{2.5}$C$_{18}$H$_{14}$ (Fig. 2(e)) did not cause the big rotation between two successive benzene rings with respect to each other. However, the arrangement of organic molecules appears more arbitrary in K$_{2.5}$C$_{18}$H$_{14}$. Comparing their XRD with the experiment one in Fig. 1, we were surprised to find that K$_{2.5}$C$_{18}$H$_{14}$ produced some peaks fitting experiment near $2\theta=10.4^{\circ}$ and $2\theta=33.2^{\circ}$ and being not observed in K$_{1.5}$C$_{18}$H$_{14}$. Of course, the XRD obtained from K$_{2.5}$C$_{18}$H$_{14}$ can not completely match that of experiment, either. Seen from the XRD result, the superconducting phase observed in the experiment is more like a mixed phase by K$_2$C$_{18}$H$_{14}$ and K$_3$C$_{18}$H$_{14}$.

\subsection{Stability}

To examine the thermodynamic stability of these considered doping cases, we have calculated the formation energy $E_f$. The $E_f$ for the doping level $x$ is defined as the function of K chemical potential the following as
\begin{equation}
E_{\text{f}}=E_{\text{doped}}-E_{\text{pristine}}-x\mu_{\text{K}}^{\text{bulk}}-x[\mu_{\text{K}}-\mu_{\text{K}}^{\text{bulk}}]
\end{equation}
where $E_{\text{doped}}$ and $E_{\text{pristine}}$ is the total energy of the doped and host crystal, respectively. $\mu_{\text{K}}^{\text{bulk}}$ can be obtained from the energy per K atom in the K metal with the $bcc$ structure. $x$ is the doping concentration. $\mu_{\text{K}}$ is the chemical potential of the K specie. $\mu_{\text{K}}=\mu_{\text{K}}^{\text{bulk}}$ means the element is so rich that the pure element phase can form. $E_{\text{f}} < 0$ indicates that the doped compound can stably exist. From the calculated formation energy shown in Fig. 3, all of considered doped phases are able to exist as the chemical potential of K satisfying certain conditions. Comparing several doping levels, however, K$_{2.5}$C$_{18}$H$_{14}$ is more stable since it has the lower formation energy than other phases in a wide range of chemical potential. This suggests that the experimentally observed superconducting phase of 7.2 K shall be K$_{2.5}$C$_{18}$H$_{14}$ or a mixture phase.

\begin{figure}
\includegraphics[width=7cm]{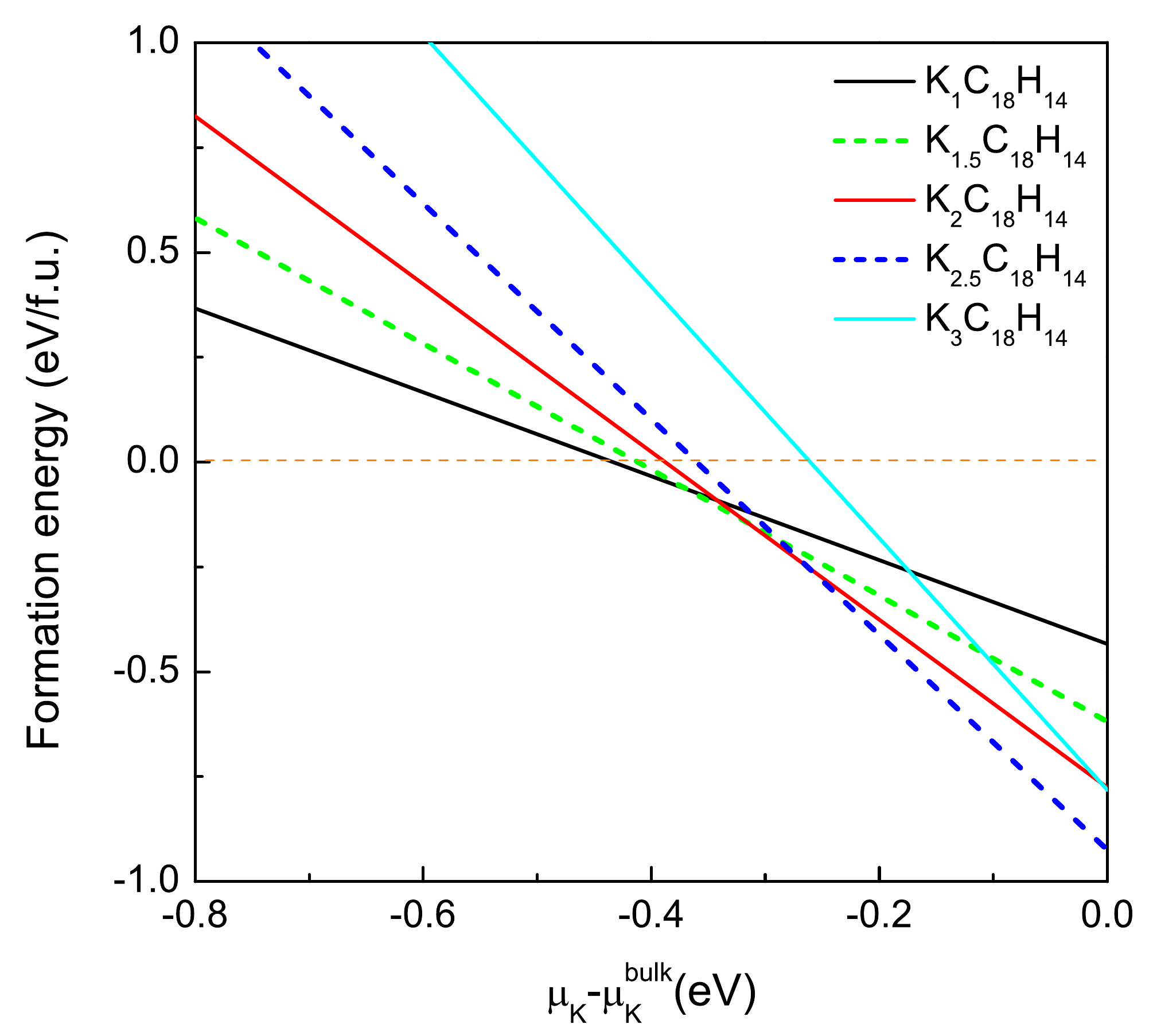}
\caption{(Color online) Calculated formation energy of K$_x$C$_{18}$H$_{14}$ as a function of the K chemical potential.}
\end{figure}

\subsection{Bonding character and charger transfer}

\begin{figure}
\includegraphics[width=\columnwidth]{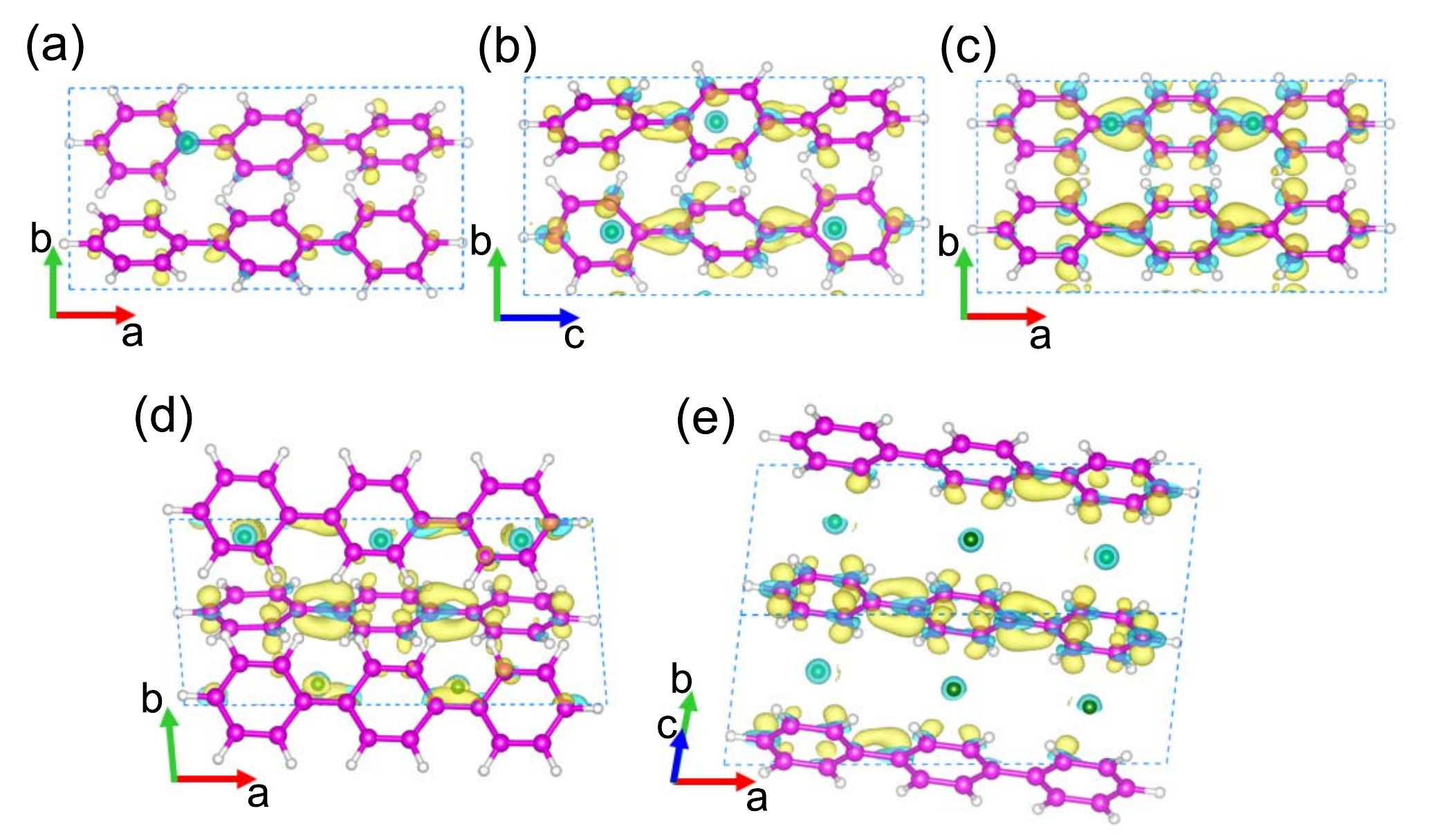}
\caption{(Color online) Calculated 3D plots of different charge density for K$_x$C$_{18}$H$_{14}$ with the iso-surface unit of $5\times10^{-3}$ e/a.u.$^{3}$. Pink, white and green balls represent C, H and K atoms, respectively. The yellow and light blue areas mean the increase and the decrease of electrons in this region, respectively. (a) K$_1$C$_{18}$H$_{14}$, (b) K$_{1.5}$C$_{18}$H$_{14}$, (c) K$_2$C$_{18}$H$_{14}$, (d) K$_{2.5}$C$_{18}$H$_{14}$ and (e) K$_3$C$_{18}$H$_{14}$.}
\end{figure}

Similar to other PAH superconductors, K-doped p-Terphenyl possesses the typical feature of charge transfer salt. Analyzing the interaction between K atoms and organic molecules, we can observe the clear ionic bonding characteristic. The different charge density $\Delta\rho$ [$\Delta\rho=\rho$(K$_x$C$_{18}$H$_{14})-\rho$(C$_{18}$H$_{14})-\rho$(K$_x$)] shown in Fig. 4 graphically depicts the charge transferring between K atoms and organic molecules. The yellow and light blue areas define as the gain and loss of electrons, respectively. With the increase of doping content, the quantity of transferred charge is about 0.84, 1.24, 1.65, 1.99 and 2.44 e/f.u. for K$_1$C$_{18}$H$_{14}$, K$_{1.5}$C$_{18}$H$_{14}$, K$_2$C$_{18}$H$_{14}$, K$_{2.5}$C$_{18}$H$_{14}$ and K$_3$C$_{18}$H$_{14}$, respectively. As mentioned above, the superconducting phase observed by experiment is predicted as a mixed phase such as the doping level around $x\sim2.5$. Then the transferring charge of about two electrons is suggested. Another important information from $\Delta\rho$ is that it visibly shows the distribution of charge transferred to C atoms. As shown in Fig. 4, the transferred charge distribution is neither local nor homogeneous. The charge is distributed on part C atoms in certain an symmetrical ordering and highlights the distribution near bridge bonds. As a comparison, the added charge mainly concentrates on the middle benzene ring with the increase of K content. To K$_3$C$_{18}$H$_{14}$, the distribution of transferred charge covers all C atoms.

\subsection{Electronic structures}

\begin{figure*}
\includegraphics[width=\textwidth]{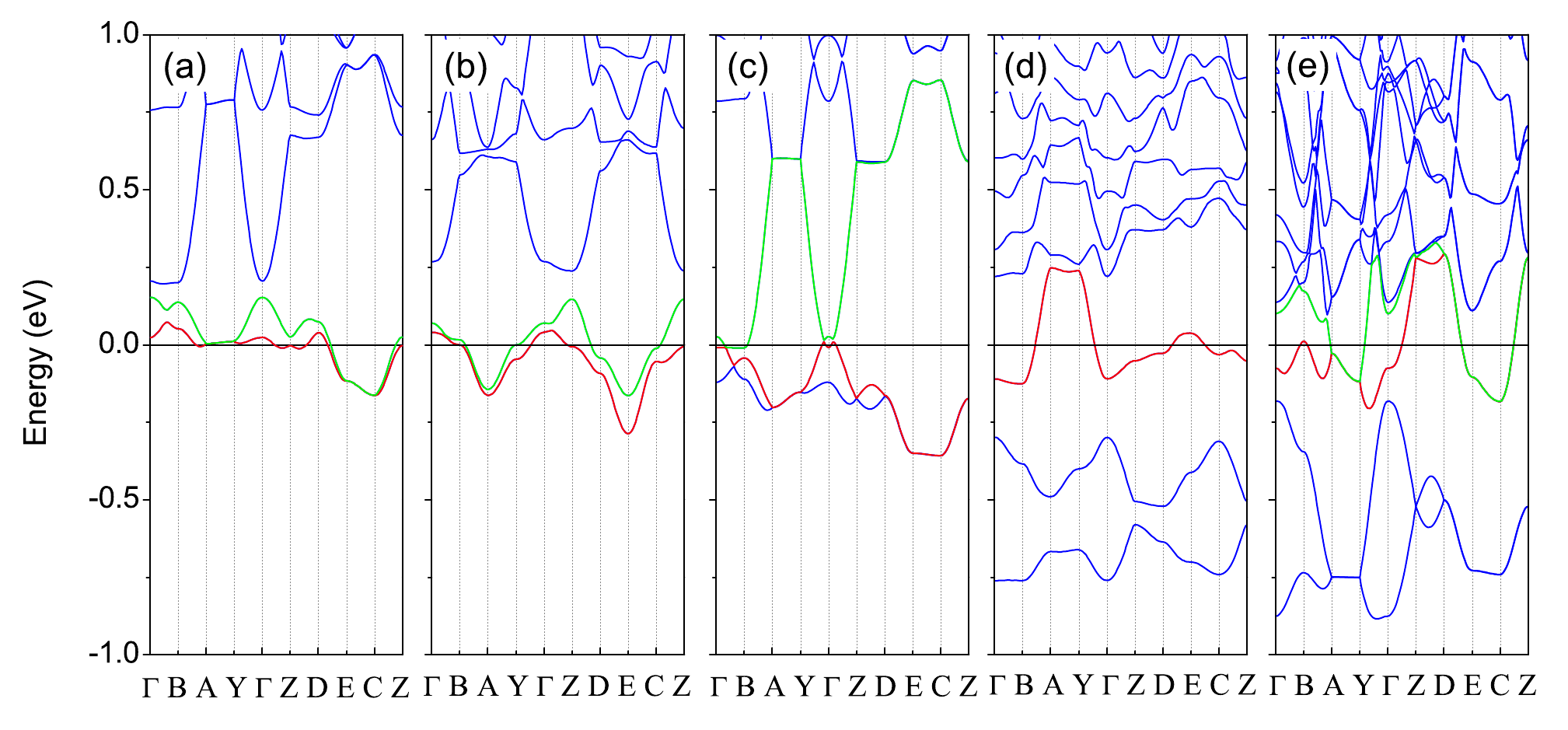}
\caption{(Color online) Electronic band structures of K$_x$C$_{18}$H$_{14}$ at NM state. Zero energy denotes the Fermi level. (a) K$_1$C$_{18}$H$_{14}$, (b) K$_{1.5}$C$_{18}$H$_{14}$, (c) K$_2$C$_{18}$H$_{14}$, (d) K$_{2.5}$C$_{18}$H$_{14}$ and (e) K$_3$C$_{18}$H$_{14}$.}
\end{figure*}

The pristine \emph{p}-terphenyl is a wide-gap semiconductor with the bandgap of 3.3 eV\cite{ref43}. The charge transferring from K atoms to organic molecules makes Fermi level shift toward to higher energy, which results in a transition from insulator to metal. Figure 5 and 6 respectively show the band structure and the total density of states (DOS) for each doping level at non-magnetic (NM) state. Under $P2_1$ symmetry, the conjugated molecules splits each orbital into a pair of partly degenerate bands in K$_1$C$_{18}$H$_{14}$. As shown in Fig. 5(a), two partly degenerate bands forming the first group conduction band cross Fermi level. However, transforming to K$_{1.5}$C$_{18}$H$_{14}$ with $P1$ symmetry (Fig. 5(b)), the coupling between the two conjugated molecules is weakened, which leads to a bigger splitting of bands. These former two doping levels both drive the high DOS at Fermi level ($N_{E_F}$) as shown in Fig. 6(a) and 6(b). The values of $N_{E_F}$ are 10.25 and 10.43 states/eV/f.u. for K$_1$C$_{18}$H$_{14}$ and K$_{1.5}$C$_{18}$H$_{14}$, respectively. More electrons transferring to organic molecules, in K$_2$C$_{18}$H$_{14}$ shown in Fig. 5(c), the Fermi level moves into the overlap region of two group of conduction bands formed by two molecular orbitals. K$_2$C$_{18}$H$_{14}$ exhibits the weak metallic feature with a small $N_{E_F}$ value of 1.28 states/eV./f.u. as shown in Fig. 6(c). In K$_{2.5}$C$_{18}$H$_{14}$ (Fig. 5(d)), more electrons transferring to organic molecules rises Fermi level into the second group of conduction band. However, the conjugated character of organic molecules completely disappears, which further weakens the coupling among bands. One band crosses Fermi level, differing from the two-band model in other doping cases. The Fermi level is localized at a DOS peak (Fig. 6(d)) which results in a slightly big $N_{E_F}$ value of 5.1 states/eV/f.u.. For K$_3$C$_{18}$H$_{14}$ with $P2_1/c$ symmetry, as shown in Fig. 5(e), the Fermi level moves into the second group of conduction band. The interactions between K atoms and molecules strengthen which makes bands become more extended. As shown in Fig. 6(e), a stronger metallic feature (3.57 states/eV/f.u. for $N_{E_F}$) is observed in K$_3$C$_{18}$H$_{14}$ than K$_2$C$_{18}$H$_{14}$. The result of electronic structures also indicates that we can obtain higher $N_{E_F}$ value by tuning the chemical potential of K to realize the high superconductivity, which is consistent with Mazziotti's suggestion\cite{ref44}.
\begin{figure}
\includegraphics[width=7cm]{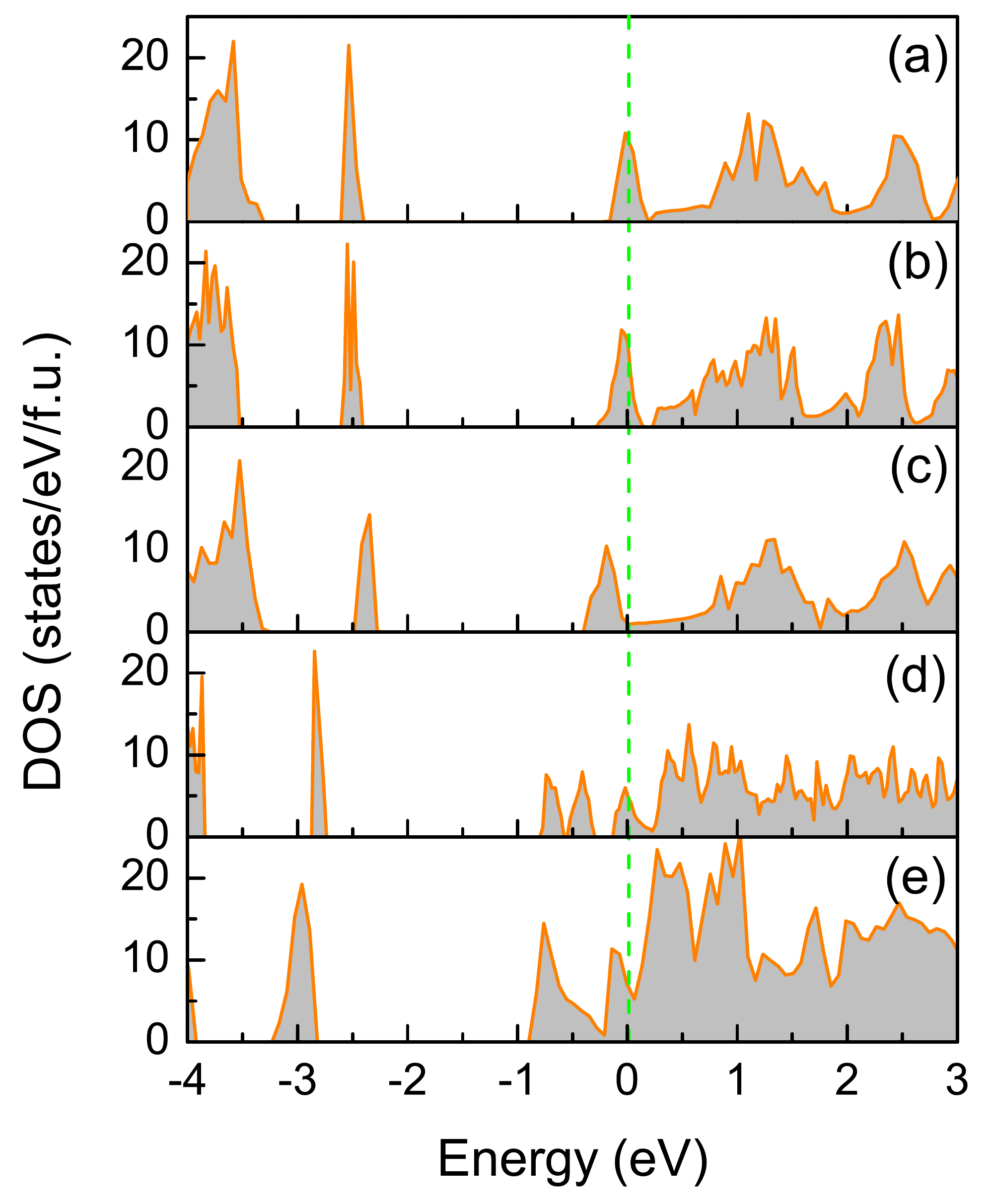}
\caption{(Color online) Electronic DOS of K$_x$C$_{18}$H$_{14}$ at NM state. Zero energy denotes the Fermi level. (a) K$_1$C$_{18}$H$_{14}$, (b) K$_{1.5}$C$_{18}$H$_{14}$, (c) K$_2$C$_{18}$H$_{14}$, (d) K$_{2.5}$C$_{18}$H$_{14}$ and (e) K$_3$C$_{18}$H$_{14}$.}
\end{figure}

Previous studies have pointed out that doped PAHs are often at the AFM ground state\cite{ref10,ref11,ref12,ref13}. Hence, we have investigated the magnetism of K-doped \emph{p}-terphenyl. It was found that both K$_1$C$_{18}$H$_{14}$ and K$_{1.5}$C$_{18}$H$_{14}$ are at the AFM ground state. The total energy of AFM state is respectively 40.4 meV for K$_1$C$_{18}$H$_{14}$ and 9.2 meV for K$_{1.5}$C$_{18}$H$_{14}$ less than that of their NM state. The local magnetic moment are respectively 0.57 and 0.22 $\mu_B$/f.u. for K$_1$C$_{18}$H$_{14}$ and K$_{1.5}$C$_{18}$H$_{14}$, respectively. The electronic states at Fermi level shown in Fig. 7 visibly decrease, and the values of $N_{E_F}$ reduce to 1.6 and 5.5 states/eV/f.u. for K$_1$C$_{18}$H$_{14}$ and K$_{1.5}$C$_{18}$H$_{14}$, respectively. Interestingly, all of K$_2$C$_{18}$H$_{14}$, K$_{2.5}$C$_{18}$H$_{14}$ and K$_3$C$_{18}$H$_{14}$ are stabilized at NM state. Comparing these magnetic structures in K-doped \emph{p}-terphenyl, the difference of magnetism possibly results from the distortion in organic molecular plane. As shown in Fig. 2, the tilt between two successive benzene rings drives the spin ordering of electrons transferred to C atoms from K atoms. With the increase of charge transferring, the spin ordering disappears in systems while the superconductivity occurs.
\begin{figure}
\includegraphics[width=7cm]{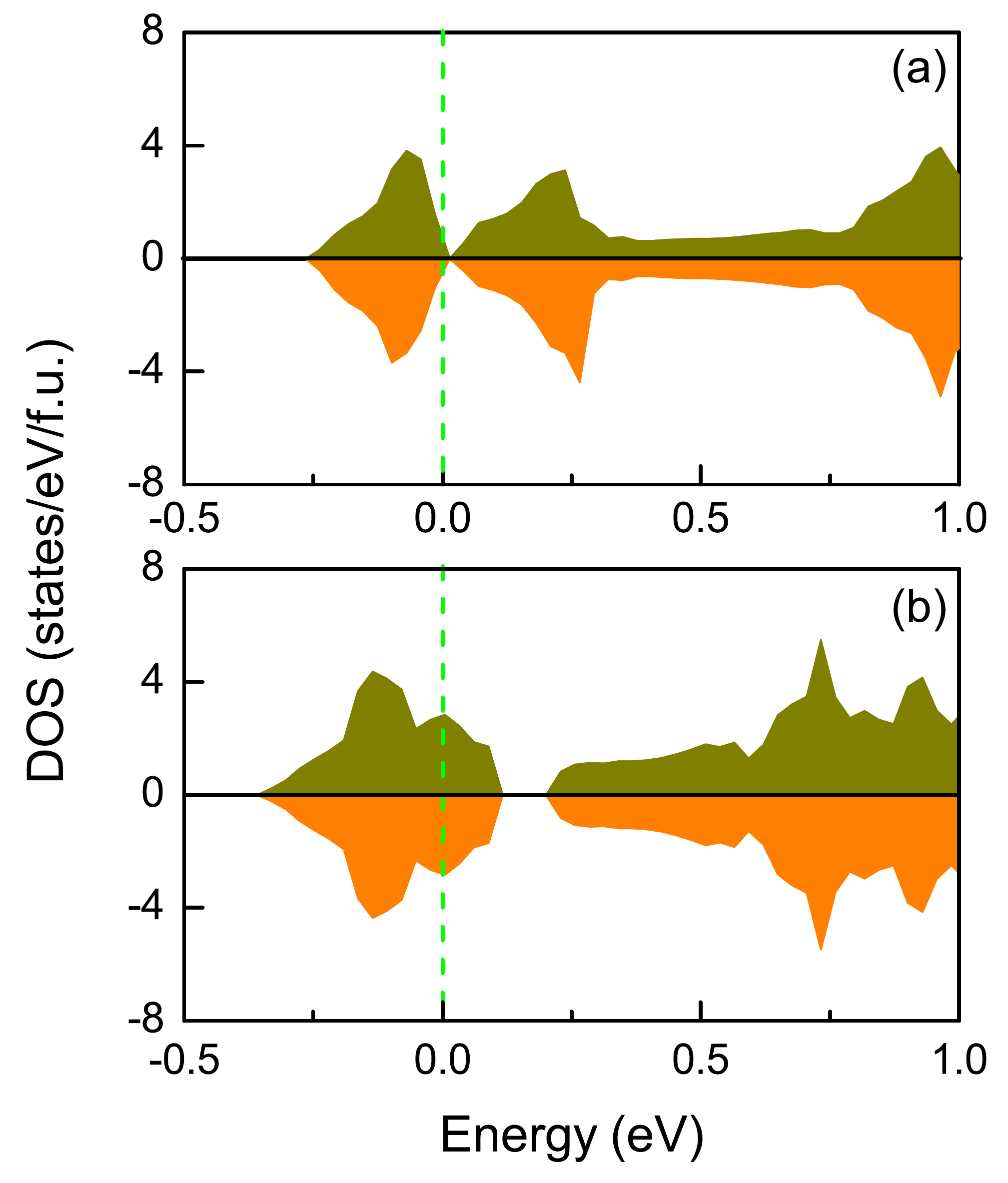}
\caption{(Color online) Electronic DOS of K$_1$C$_{18}$H$_{14}$ (b) and K$_{1.5}$C$_{18}$H$_{14}$ (b) with the AFM spin polarization. Zero energy denotes the Fermi level.}
\end{figure}

\subsection{Discussions}

With regard to the superconductivity of K-doped \emph{p}-terphenyl, more data and analysis are yet to come. The complexity of superconducting mechanism in K-doped \emph{p}-terphenyl is evidented by recent reported three critical temperatures, 7.2, 43 and 123 Kelvin, in the same material\cite{ref22,ref23,ref24}. In previous studies, however, we have pointed out that there is a common phase to show $T_c$ in the range of $5<T_c<7$ Kelvin in all K-doped aromatic compounds\cite{ref9}. The existed superconductivity of 7.2 Kelvin in K-doped \emph{p}-terphenyl just confirms that prediction. For this low $T_c$ phase, the electron-phonon coupling mechanism is enough to describe this superconductivity. From previous investigations\cite{ref9,ref45}, the phonons with low and middle frequency mainly contribute to electron-phonon interaction. The maximum value of our calculated the middle frequency of K$_{2}$C$_{18}$H$_{14}$ is about 1574 cm$^{-1}$ which is almost equal to that of K-doped picene\cite{ref45} and also comparable with that of K$_2$C$_6$H$_6$\cite{ref9}. At the same time, the $N_{E_F}$ can change with the doping content around 5.1 states/eV/f.u. for K$_{2+\delta}$C$_{18}$H$_{14}$ ($0<\delta<1$) due to the charge fluctuations. Based on the comparable $N_{E_F}$ of K-doped \emph{p}-terphenyl with K-doped benzene and picene in the range of $4-6$ states/eV/f.u.\cite{ref9,ref45}, and combining with the similar feasible screened Coulomb pseudopotential $\mu^{\star}$ of 0.1, we can predict that the $T_c$ of K-doped \emph{p}-terphenyl is in the range of $5<T_c<7$ Kelvin, which is according with the experimental observation\cite{ref23}. In other words, the observed superconducting phase is a mixed compound in the doping range of $2< x<3$ instead of pure K$_1$C$_{18}$H$_{14}$ of K$_{1.5}$C$_{18}$H$_{14}$ phase, which is just what we predicted by XRD, thermodynamic stability, electronic and magnetic properties.

With regarded to higher transition temperature, the experimental investigations presented unknown peaks in Raman spectra. These unknown peaks are possibly induced by graphite, C$_{60}$ and other modes\cite{ref23,ref24}. On the one hand, therefore, we infer that the two higher transition temperatures may be from other new superconducting modes such as CK$_x$, KH$_x$ and K-CH$_x$ compounds induced by the reconstruction. On the other hand, as our previous studies\cite{ref9,ref19}, some internal and external factors can be the cause of high superconducting transition temperature, such as electronic correlations and pressure. The strong electronic correlations in K-doped p-terphenyl were implied by Baskaran\cite{ref46} and Fabrizio \emph{et al.}\cite{ref47}. In this kind of low dimensional organic system, therefore, the high superconductivity may be related with the electronic correlations effects. To clarify the matter, more efforts in the future are surely needed.

\section{Conclusions}
In conclusion, with the aim of exploring the structural and bonding characteristics of K-doped \emph{p}-terphenyl which has been discovered to be a superconductor with $T_c=7.2-123$ Kelvin, we have carried out the first-principles calculations based on vdW-DF2 functional. Considering five doping levels of $x=$ 1, 1.5, 2, 2.5 and 3 in K$_x$C$_{18}$H$_{14}$, we have predicted the optimized crystal structures at each doping level, calculated the XRD spectra, and investigated the thermodynamic stability, ionic bonding characteristics, charge transfer, electronic and magnetic properties. All of these five doping phases are able to exist in experiment based on the negative formation energy, but K$_{2.5}$C$_{18}$H$_{14}$ or say a mixed phase by K$_2$C$_{18}$H$_{14}$ and K$_3$C$_{18}$H$_{14}$ is more stable in a wide range of chemical potential. The XRD summarized by K$_2$C$_{18}$H$_{14}$, K$_{2.5}$C$_{18}$H$_{14}$ and K$_3$C$_{18}$H$_{14}$ can almost match that of experiment. The charge transfer from K atoms to organic molecules results in the insulator-metal transition. However, both K$_1$C$_{18}$H$_{14}$ and K$_{1.5}$C$_{18}$H$_{14}$ are stabilized at AFM ground state, while the latter three compounds exhibit the non-magnetic behavior. The superconducting phase observed in experiment should be a mixture phase with doping level in range of $2<x<3$. However, the multiformity of band structures near Fermi level indicates that the driving force for superconductivity is complicated. Future works are needed to understand the superconductivity especial for the transition temperature above 40 Kelvin.

\section{ACKNOWLEDGMENTS}
The work was supported by the National Natural Science Foundation of China (Grant Nos. 61574157 and 61774164) and the Basic Research Program of Shenzhen (Grant nos. JCYJ20160331193059332, JCYJ20150529143500956 and JCYJ20150401145529035). H. Q. Lin, X. H. Wang, J. X. Han and G. H. Zhong acknowledge support from NSAF U1530401 and computational resource from the Beijing Computational Science Research Center. The partial calculation was supported by the Special Program for Applied Research on Super Computation of the NSFC-Guangdong Joint Fund (the second phase) under Grant No. U1501501.

\bibliography{aipsamp}

\end{document}